\documentclass[prd,aps,showpacs,superscriptaddress,twocolumn]{revtex4}

\usepackage{graphicx}
\def\mev{{\rm MeV}}

\def\fm{\text{ fm}}
 
\def\csw{c_{\rm sw}}

\newcommand{\be}{\begin{equation}}
\newcommand{\ee}{\end{equation}}
\newcommand{\bea}{\begin{eqnarray}}
\newcommand{\eea}{\end{eqnarray}}

\newcommand{\eq}[1]{eq.\,(\ref{#1})}
\newcommand{\fig}[1]{Fig.\,\ref{#1}}
\newcommand{\tab}[1]{Table\,\ref{#1}}

\newcommand{\figurebox}[2]{\fbox{\vbox to#2in{\hbox to #1in{\hfil}\vfil}}}
\def\lsi{\raise0.3ex\hbox{$<$\kern-0.75em\raise-1.1ex\hbox{$\sim$}}}
\def\gsi{\raise0.3ex\hbox{$>$\kern-0.75em\raise-1.1ex\hbox{$\sim$}}}

\newcommand{\MeV}{\mathop{\rm MeV}}

\newcommand{\swansea}{
Department of Physics, University of Wales Swansea, Swansea
SA2~8PP, Wales}

\newcommand{\edinburgh}{
School of Physics, University of Edinburgh,
Edinburgh EH9~3JZ, Scotland}

\newcommand{\liverpool}{Division of Theoretical Physics, 
Department of Mathematical
Sciences, University of Liverpool, Liverpool L69~7ZL, England}

\newcommand{\figwidth}{2in}

\begin{document}

\title{Improved Wilson QCD simulations with light quark masses}

\author{C.R. \surname{Allton}} 
\affiliation{\swansea}

\author{A. \surname{Hart}}
\author{D. \surname{Hepburn}} 
\affiliation{\edinburgh} 

\author{A.C. \surname{Irving}}
\affiliation{\liverpool}

\author{B. \surname{Jo\'o}}
\affiliation{\edinburgh}

\author{C. \surname{McNeile}}
\author{C. \surname{Michael}}
\author{S.V. \surname{Wright}}
\affiliation{\liverpool}

\collaboration{UKQCD Collaboration}
\noaffiliation

\begin{abstract}
  We present results from simulations using 2 flavours of ${\cal
  O}(a)$-improved Wilson quarks whose masses are about $1/3$ of the
  physical strange quark mass.  We present new data on the mass of the
  singlet pseudoscalar meson and evidence of the onset of chiral
  logarithms in the pion decay constant.  The previously observed
  suppression of the topological susceptibility at lighter quark masses
  is confirmed. We report on the performance of the hybrid Monte Carlo
  algorithm at light quark masses.
\end{abstract}

\preprint{Edinburgh 2004/05}
\preprint{Liverpool Preprint LTH619}
\preprint{Swansea Preprint SWAT/xxx}


\maketitle

  \section{Introduction}
  \label{sec:intro}
  The first generation of QCD simulations using Wilson fermions have
  provided useful information on the hadron spectrum but have
  been restricted to relatively heavy quark masses 
  (see for example the annual reviews by Aoki~\cite{Aoki:2000kp}
  and Kaneko~\cite{Kaneko:2001ux}). 
  Using an improved staggered discretisation, 
  the MILC collaboration~\cite{Bernard:2001av}
  has probed the spectrum with significantly lighter 
  quark masses and very promising results are now being
  obtained~\cite{Davies:2003ik} from simulations with 
  light quark masses down to $m_s/8$, a physical strange quark mass
  $m_s$ and lattice spacings down to $0.09$~fm.
  There are theoretical and numerical complications associated with 
  this action and the representation of lattice flavour symmetries
  and so simulations with Wilson fermions remain an important
  tool in studying full QCD. In due course, simulations with chiral
  fermions should become feasible and provide a further  
  cross-check on the results now being obtained.

  In an earlier paper~\cite{Allton:2001sk} the UKQCD collaboration has
  presented lattice QCD results based on two-flavour simulations
  conducted at fixed lattice spacing ($a\approx 0.1$ fm) and fixed
  volume around $(1.6 \fm)^3$.  This work also contained results
  obtained at fixed gauge coupling ($\beta=5.2$) also using the standard
  Wilson gauge action and ${\cal O}(a)$--improved Wilson fermions. The
  lightest sea quark mass achieved in these simulations was around $0.6$
  of the strange quark mass ($m_{\pi}/m_{\rho}\approx 0.58$).  There is
  evidence that some hadronic observables suffer finite size effects
  already on these lattice volumes~\cite{Aoki:2002uc,AliKhan:2002hz}.
  Working at lighter quark masses at fixed gauge coupling is expected to
  enhance further these effects and to provide an even greater challenge
  to the standard Monte Carlo simulation algorithms.  In this paper, we
  report on the results of an attempt to push our analysis towards this
  lighter quark regime in order to (a) uncover any more obvious effects
  of dynamical fermions not hitherto seen with this
  action~\cite{Allton:2001sk,Aoki:2002uc} and (b) determine the limits
  if any of the simulation algorithm in its simplest form. A previous
  study~\cite{Joo:2000dh} had indicated a potential instability at light
  quark masses when using step sizes that were too large in the
  molecular dynamics trajectory part of the update.

  This paper should be seen as an addendum to our previous paper in which
  full details are given of the action, simulation methods and
  basic hadronic measurements. Some preliminary results from this 
  work were presented in~\cite{Irving:2002fx,Hepburn:2002wa}.

  Recent related work using improved Wilson fermions includes that of
  the JLQCD collaboration~\cite{Aoki:2002uc} which uses the same action and
  covers a similar range of quark masses to our earlier work but which also 
  includes simulations in larger volumes ($\approx (1.8\,\hbox{fm})^3$). 
  The QCDSF collaboration has been conducting simulations complementary to those
  of UKQCD, most recently using an improved algorithm~\cite{AliKhan:2003br}.
  Neither of these simulations sets has penetrated the lighter quark regime.
  In each case $m_{\pi}/m_{\rho}\geq 0.6$. 

  The qq+q collaboration~\cite{Farchioni:2003bx} has succeeded in 
  simulating at one half (or less) of the strange quark mass 
  but with rather coarse lattice spacing ($a\approx 0.28$ fm).

  The rest of the paper is as follows.  In section~\ref{sec:simul}, we
  describe the parameters and performance of the new simulations.  The
  lattice spacing is determined from the static potential and
  decorrelation effects are studied using measurements of the
  topological susceptibility.  In section~\ref{sec:spect}, we present
  the additions to our previous collection of data for the hadron
  spectrum and meson decay constants and use them to search for first
  signs of chiral logs.  Results from new measurements of disconnected
  loops, including the $\eta$ mass, are presented in
  section~\ref{sec:discon}.  We present results on the topological
  susceptibility in section~\ref{sec:top}.
  Conclusions are drawn in section~\ref{sec:conc}.

  %
  %
  \section{Simulations with improved Wilson fermions}
  \label{sec:simul}

\begin{figure}[t]
  \begin{center}
  \leavevmode
  \includegraphics[width=\figwidth,clip]{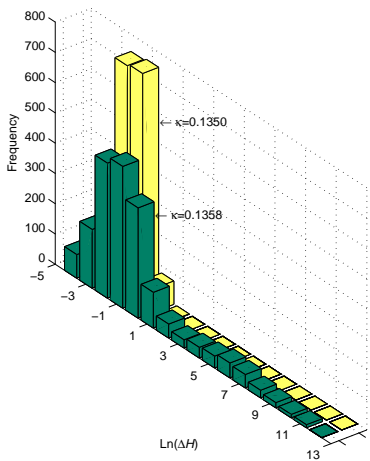}
  \end{center}
  \caption[]{\label{fig:histldh} {Histogram of $\ln\Delta H$
  (positive values of $\Delta H$ only) for this ensemble 
  (darker bars) and for a similar sized sample from
  simulations at $\kappa=0.1350$.
  }}
  \end{figure}

  The simulations were conducted using the standard Hybrid Monte Carlo
  (HMC) algorithm as described in~\cite{Allton:2001sk} using lattice
  action parameters $(\beta,\csw,\kappa)=(5.2,2.0171,0.1358)$
  and a lattice volume $V =
  L^3T = 16^3 32$.  A total of $2440$ trajectories were accumulated at
  the rate of about $3$ per hour on a machine sustaining $30$ Gflops.
  In physical units, the lattice volume was $L^3\approx (1.5 \fm)^3$ and
  $m_\pi\approx 420 \MeV$ before chiral or continuum extrapolation.  The
  lattice spacing was estimated from the measured hadronic scale
  parameter
  \cite{Sommer:1994ce}:
  \begin{equation}
  r_0/a = 5.32(5) \, .
  \label{eq:r0} 
  \end{equation}
  The integrated autocorrelation time $\tau$ of the mean plaquette was
  found to be $6.9(14)$ trajectories. This follows the previously
  observed trend of a {\em decrease} of $\tau$ with decreasing quark
  mass.  For comparison, $\tau=16(3)$ at $\kappa=0.1350$ where
  $m_{\pi}/m_{\rho}\approx 0.7$. 
  For the
  present ensemble, $m_{PS}/m_V=0.44(2)$ (see later).  Although not
  expected, this trend can be accommodated in simple
  models~\cite{Allton:2001sk}.  The mean plaquette was
  \begin{equation}
  \left\langle P \right\rangle = 0.53767 (3) \, .
  \label{eq:plaq}
  \end{equation}
  These HMC runs are expected to be susceptible to instabilities
  (occasional large $\Delta H$ values and zero acceptance) when the
  fermion force term gets too large
  \cite{Joo:2000dh}.
  We have observed this effect directly in the present simulations where
  we found it necessary to use a step size of $1/400$.  When the step
  size is this, or smaller, we found that $64$ bit arithmetic for field
  storage and matrix-vector manipulations was required so as to avoid a
  serious loss of acceptance due to rounding errors (see below).  This
  was true even though we always used full 64-bit arithmetic and careful
  summing techniques for the global summation in our $\Delta H$ (energy
  difference) calculations.  \fig{fig:histldh} shows a histogram
  of $\ln \Delta H$ for all trajectories where $\Delta H>0$. For
  comparison we also show the corresponding histogram for the
  well-behaved simulation at $(\beta,\kappa)=(5.2,0.1350)$ described in
  \cite{Allton:2001sk}.  
  The sample size in each case is the same.

  The anomalous trajectories leading to very large values of $\Delta H$
  are clearly visible in the simulation at $\kappa=0.1358$.  The
  simulation was an experimental one and incorporated several changes of
  simulation parameters (step size and solver accuracy \textit{etc.})
  leading to large changes in acceptance.  Thus one should bear in mind
  the possible consequences of this on the discussion of autocorrelation
  times and error analysis.

\begin{figure}[t]
  \begin{center}
  \leavevmode
  \includegraphics[width=\figwidth,angle=-90,clip]{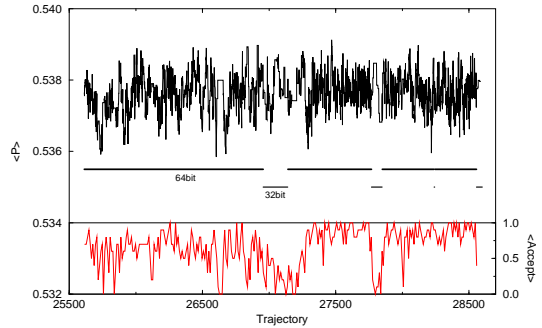}
  \end{center}
  \caption[]{\label{fig:history} {The time history of the mean plaquette
  together with the corresponding HMC acceptance (locally averaged
  over 10 trajectories). The horizontal bars indicate the precision used as 
  discussed in the text.
  }}
  \end{figure}

  In \fig{fig:history} we show a time history of the mean plaquette
  along with the average acceptance (integrated over 10 trajectories).
  The dramatic drop in acceptance associated with a change to 32-bit
  arithmetic is clearly visible. The location
  of the changes are indicated by the horizontal bars.  The loss of
  acceptance was not significantly dependent on the size of the solver
  residual in the molecular dynamics steps (using a running rather than
  absolute residual).  As noted above, it was primarily dependent on the
  arithmetic used in the matrix-vector calculations.

  Despite the rather chequered history of the configurations we decided
  to subject them to physics analysis.  This ensemble represented an
  expensive investment in computer time and promised to give access to
  relatively light quark masses (by Wilson lattice quark standards).
  In view of the above remarks, one should bear in mind the deficiencies
  of the Markov process that led to their generation.  We may attempt, in due
  course, to repeat the simulations at this lattice spacing and quark
  mass using improved algorithms, faster machines and larger lattices.

  \section{Hadron spectrum and decay constants}
  \label{sec:spect}

   In this section we report on the light spectroscopy from this ensemble.
  We use a similar analysis to our original work on the spectroscopy of
  nonperturbatively-improved clover fermions  at $\beta$ = 5.2~\cite{Allton:2001sk}. Here
  we concentrate on the unitary  sector of the theory --- with valence
  quarks equal in mass to the sea quarks.
  We have previously reported~\cite{Irving:2001vy}
  some evidence for chiral logs
  in a partially quenched analysis with $\kappa_{\mathrm{sea}}$ = 0.1355 and
  $\kappa_{\mathrm{valence}}$ = 0.1358.
   We use fuzzed and local sources and sinks combined to make a
  variational fit. To determine the pion decay constant, we fit with two
  states an order 4 variational matrix with fuzzed and local sources
  using the $\gamma_5$ and $\gamma_4 \gamma_5$ operators to create a pion.

  To increase statistics, we use quark propagators  with sources on the
  time planes 7, 15, and 23 in addition to the $t=0$ plane for $\kappa$
  values 0.1358, 0.1355, and 0.1350. In our first published work we  only
  used the quark propagators from time plane $t=0$. Thus we report new numbers
  at $\kappa$ = 0.1350 and 0.1355 for  $\pi$ and $\rho$ with reduced error
  bars.

\begin{figure}
  \begin{center}
  \includegraphics[scale=0.35]{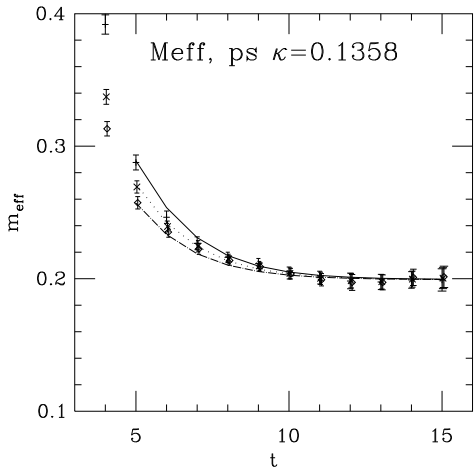}
  \caption[]{Effective mass plot for the pseudoscalar channel
  at $\kappa = 0.1358$.}
  \label{fg:Meff}
  \end{center}
  \end{figure}

  In \fig{fg:Meff} we show the effective mass plot for the pseudoscalar
  channel (PS) at $\kappa = 0.1358$. The results from the fits are in
  Tables~\ref{tab:conData}, \ref{tab:fpiData}. We investigated the
  stability of the fits in a number of ways.  Since the spectrum fits used
  two states, the ground state is expected (and indeed found) to be very
  stable while the excited state values should be regarded as indicative,
  although these values  are consistent with the expected lightest
  multi-body states at $3 m_{\pi}$ and $2 m_{\pi}(k=2\pi/L)$, for
  pseudoscalar  and vector, respectively.

  \begin{table}
  \begin{center}
  \begin{tabular}{|lllll|} \hline
  $\kappa$ & Hadron & $t-$range &  $a m_0$  & $a m_1$ 
  \\  \hline
  $0.1358$ & $0^{-+}$ & 4 - 15 &
  $0.199(5)$ & $0.75(11)$  \\
  $0.1358$ & $1^{--}$ & 4 - 12 &
  $0.450(14)$ & $1.16(6)$ \\
  $0.1355$ & $0^{-+}$ & 4 - 15 &
  $0.282(4)$ & $ 0.90(11)$  \\
  $0.1355$ & $1^{--}$ & 4 - 12 &
  $0.491(7)$ & $1.17(3)$ \\
  %
  $0.1350$ & $0^{-+}$ & 4 - 15 &
  $0.408(2)$ & $1.19(8)$  \\
  $0.1350$ & $1^{--}$ & 4 - 12 &
  $0.585(4)$ & $1.33(3)$ 
  \\ \hline
  \end{tabular}
  \end{center}
   \caption{
  Masses in lattice units from this calculation.
  }
   \label{tab:conData}
  \end{table}

  \begin{table}
  \begin{center}
  \begin{tabular}{|lll|} \hline
  $\kappa$ & $af_A$ &  $af_P$ \\  \hline
  0.1358 & 0.0829(26) &  0.1457(78) \\
  0.1355 &  0.1055(14) &  0.1835(44) \\
  0.1350 & 0.1336(11) &  0.2468(33) \\
   \hline
  \end{tabular}
  \end{center}
   \caption{ The raw lattice value of $af_{\pi}$ is  given by using the
  order $a$ improved expression $(1+b_A m)(af_A+c_A af_P)$ and we tabulate
  these two contributions.
  }
   \label{tab:fpiData}
  \end{table}

  Our values of $m_{\pi} L \approx 4$, so we should expect some finite
  volume  effects. Using a similar formalism and parameters (the only
  difference being to  use $c_{SW}=2.02$ rather than 2.017), 
  JLQCD~\cite{Aoki:2002uc}  have explored this for the hadron spectrum
  using $L=12,\ 16, \ 20$. Here we discuss this at  their lightest
  $\kappa$-value (0.1355), where they do  see evidence of a finite size
  effect (masses lower at larger volume) although it is not very
  significant statistically for the two larger volumes. We have a larger
  number of trajectories at $L=16$ than JLQCD,  and there are some
  statistically significant differences between our 
  results~\cite{Allton:2001sk} and theirs for that volume. However, for
  the  pseudoscalar and vector mesons, the differences between our new
  determinations at $L=16$ (given above) and theirs  at $\kappa=0.1355$
  are not statistically significant. Using our newer results to make the
  comparison with $L=20$ and $L=16$, then  suggests that there is a decrease
  of the pion mass  as the volume increases by about $4\pm 2 \% $. 

  %
  %
   We have also computed the pion decay constant for this data set. It has
  been clear from the work of the  GF11 group that the unquenching effects
  are larger in decay constants than for masses~\cite{Butler:1994zx}, 
  although the systematic errors on decay constants can be large due to
  truncation of  perturbative series. The ratio $f_{K}/f_{\pi}$  (for which
  renormalisation factors should largely cancel) is underestimated in
  quenched lattice QCD~\cite{Heitger:2000ay}.

  A critical goal of lattice gauge theory calculations is to detect the
  presence of chiral logs in observables. The loop corrections from chiral
  perturbation are non-analytic in the parameters of the lowest  order
  chiral Lagrangian,  and hence provide a good check that the lattice 
  calculation is in the regime where chiral perturbation theory is valid. 
  A particularly appropriate way to look for these chiral logarithms is 
  in the pseudoscalar decay constant.

  We are working, because of computational constraints, at finite  lattice
  spacing. The formalism of chiral perturbation theory can be  extended to
  cover this
  case~\cite{Aoki:2003yv,Bar:2003mh},
  but at the cost of 
  additional parameters. We choose instead to make a comparison with the 
  continuum predictions of chiral perturbation theory.

\begin{figure}
  \begin{center}
  \includegraphics[scale=0.35,angle=90]{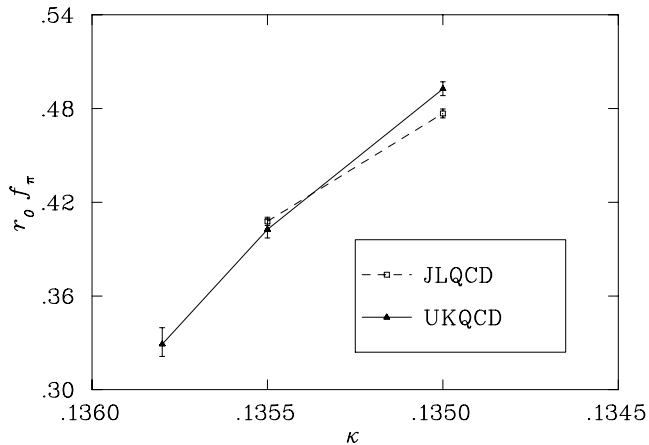}
  \caption[]{
   The pseudoscalar decay constant in units of $r_0$ from  UKQCD  and
  JLQCD versus $\kappa$.
   }
  \label{figukvsjl}
  \end{center}
  \end{figure}

   Our results for $f_{\pi}$ are shown in \fig{figukvsjl}. We have
  extracted the values  from fits as discussed above and then applied the
  rotations and corrections  appropriate for an order $a$ improved
  formalism. We have used the  same perturbative formulation of these
  corrections (and also the same prescription for  $Z_A$ ) as employed by
  JLQCD, in order to facilitate comparisons. Moreover,  since they use a
  different prescription for evaluating $r_0$, we  have applied  our
  determination of $r_0$ to their data. The comparison is shown in
  \fig{figukvsjl}.  The agreement at our two common  $\kappa$ values
  is adequate and we present here a new determination at a lighter
  $\kappa$ value. The significant feature of this new result is that  it
  shows a curvature versus $m_q$.

   This curvature may be the first evidence in this study of the chiral 
  logarithm at work. To explore this we compare our result with some 
  continuum chiral models for which the chiral logarithm has a fixed 
  coefficient (given by $f_{\pi}$). For $N_f=2$ flavours 
  of degenerate quark, the standard chiral perturbation theory 
  result to one loop is:

    \begin{equation}
      {f_{\pi}(m) \over  f_{\pi}(0)} = 1 -
   2\left( { m  \over 4 \pi f_{\pi}(0) } \right)^2
   \log ({m^2 \over\Lambda^2}) + O(m^4)
   \end{equation}

\begin{figure}
  \begin{center}
  \includegraphics[scale=0.35,clip]{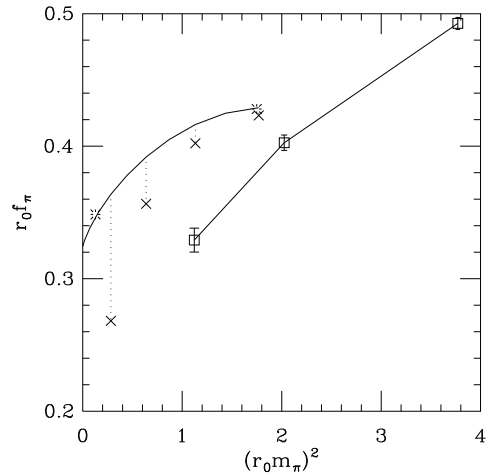}
  \caption[]{
   The pseudoscalar decay constant in units of $r_0$ from  UKQCD  
  versus the squared pseudoscalar meson mass.
   Also shown is an expression including chiral perturbation theory terms 
  to order $m^4$ which has been fitted (see ref.~\cite{Colangelo:2003hf}
  where we  use $\mu=0.75$ GeV and $\tilde{r}_F(\mu)=-2$) to agree with  the
  experimental values of $f_{\pi}$ and $f_K$ which are shown (*). An
  estimate~\cite{Colangelo:2003hf}  of the finite size effect expected from
  chiral perturbation theory (to order $m^2$) is shown by the vertical
  lines.
   }
  \label{figchlog}
  \end{center}
  \end{figure}

  This expression has an unsatisfactory behaviour at large $m$, where 
  chiral perturbation theory should not apply anyway. Moreover  even at
  the K mass, the quartic terms in a chiral perturbative treatment are 
  significant~\cite{Colangelo:2003hf}. We can thus either  concentrate on
  the  curvature implied at small $m$ or modify the expression
  phenomenologically. We illustrate this behaviour in \fig{figchlog}
  by  using an empirical determination~\cite{Colangelo:2003hf} of the  terms
  arising in chiral perturbation theory up to $m^4$. This shows the
  curvature to be expected  at small pseudoscalar mass in a large volume.
  Unfortunately the overlap between our data and the region of validity of
  the  chiral approach to this order is small.

   As discussed above, finite volume effects should be important, since for
  this lightest  quark mass, we have $m_{\pi}L=3.2$.  As has long been
  known, continuum chiral perturbation theory  in a finite
  volume gives explicit predictions. This has been explored
  theoretically~\cite{Colangelo:2003hf} using a range of different treatments
  of  the chiral formalism, showing that we are in the region where chiral
  models do give rise to significant effects. Moreover the finite size
  effects arise from chiral loops and so are from the same source as the
  logarithmic corrections. Using $L=1.5$ fm and $m_{\pi}=400$ MeV, which are
  close to our values,  a range of different levels of approximation 
  yields~\cite{Colangelo:2003hf} a relative increase of the finite-volume
  pion mass over the  infinite volume mass of 0.6 to 2\%.  This can be
  compared to the shift of $4 \pm 2\%$ obtained from lattice comparisons,
  as discussed above.  This is not a statistically significant comparison,
  but it  does indicate that the magnitude of finite size effects to be
  expected theoretically is consistent with that seen on the lattice.

   In the Gasser-Leutwyler chiral approach, the relative finite size
  effects for $f_{\pi}$ will be  four times as large as those for the pion
  mass, and of opposite  sign. Again, this finite size effect  comes from
  the one loop term that gives the chiral logarithm.   We show  the effect
  of this shift for the Gasser-Leutwyler
  approach~\cite{Bijnens:1996yn,Bijnens:1997vq,Colangelo:2003hf,Bijnens:1998fm} 
  (corresponding to  3.4\% for the
  above values  of $L=1.5$ fm and $m=400$ MeV) in \fig{figchlog}. 
  Note that
  the curvature we observe is equivalent to a decrease  of $f_{\pi}$ at
  $\kappa=0.1358$ of about 8\%, which could thus be  ascribed entirely to
  finite size effects only if the one loop estimate was less than 50 \% of
  the total. Since the finite size estimates come from chiral
  models, they will  also include a chiral logarithm which will yield 
  curvature.

  We conclude that the finite volume effects will enhance the curvature 
  at small pion masses, as indeed we find. Hence our results are in 
  qualitative agreement with chiral perturbation theory.

   We note that our data suggests an extrapolation to the chiral limit 
  which would give a value for $f_{\pi}$ below the experimental value  of
  131 MeV. Since the perturbative correction to  $Z_A$ is of order 25\%
  at first order, we expect possible systematic errors  of up to 5\% from
  the next order, which is only estimated by using tadpole-improved 
  methods. This error budget for $Z_A$ is confirmed by results from 
  quenched studies where the non-perturbative  evaluation of $Z_A$ gave a
  value 4\% different from the tadpole-improved  one loop result used
  here. We chose to use a value of $r_0=0.525$ fm to set the scale 
  following previous work ~\cite{Dougall:2003hv,Allton:2001sk} and this value is 
  uncertain to 5\%. 
   There are further errors in the lattice determinations coming from having
  a sea quark mass which is too large, from
  neglect of  the strange sea and from finite $a$
  effects (as well as the finite volume effects we have discussed  above).
  This covers the apparent discrepancy of 10\%  seen in the figure
  comparing  our result to the experimental values of  $f_{\pi}$ and
  $f_K$.

  Although it may appear that reproducing $f_{\pi}$ and $f_K$ from 
  lattice QCD has no  immediate experimental impact, that is not 
  quite true.
  The search for chiral logs in decay constants is currently one of 
  the most important topics in heavy-light 
  physics~\cite{Yamada:2002wh,Kronfeld:2003sd}. 
  The error on the ratio of the $f_{B_s}/f_B$ has recently been
  increased, because the chiral log term has not been observed
  in lattice data~\cite{Kronfeld:2002ab}.
  The ratio $f_{B_s}/f_B$ is an important QCD quantity 
  for the unitarity checks of the CKM matrix. It will become more 
  important once $B_s$ mixing is measured at run II of CDF.
  As has been noted by many 
  authors~\cite{Bernard:2002pc,Becirevic:2002mh},
  the chiral log structure 
  of $f_\pi$  and $f_B$ are rather similar. Hence, a detection of 
  chiral logs in $f_\pi$ is an indication that the parameters of the
  unquenched calculation are close to where chiral logs may occur 
  in the heavy-light decay constant.

    Evidence for chiral logs in both heavy and light QCD has been claimed
  in unquenched calculations with improved staggered
  quarks~\cite{Wingate:2003gm}. Particularly because of the complexity of
  the chiral perturbation theory calculations for  staggered
  fermions~\cite{Aubin:2003mg,Aubin:2003uc}, we feel a cross check on the
  improved staggered calculations is very valuable, even if ultimately
  unquenched clover fermions do not allow us to control all the systematic
  errors such as lattice spacing dependence.

  \section{The mass of the singlet pseudoscalar}
  \label{sec:discon}

  The large splitting between the mass of the $\eta'$ and the octet of
  light pseudoscalars is thought to be caused by the complex vacuum of QCD
  and  the anomaly in the axial current. There is a lot of  activity in
  trying to reconcile the mechanism behind the mass splitting between the
  mass of the $\eta'$ and the  masses of the octet. In particular the
  questions raised by Witten about the consistency of the  solution of the
  $U(1)$ via instantons with the  large $N_c$ limit~\cite{Witten:1979bc}
  is topical. There are also many phenomenological puzzles in which the $\eta'$
  is involved. See the review by Bass~\cite{Bass:2001ix} for a  review
  of experiments with $\eta$ and $\eta'$ as decay products. A first
  principles calculation  of the structure of the $\eta'$ would be helpful.

   In the real world the mass of the $\eta'$ is   also determined by the
  mixing between the  singlet and octet mesons. We introduce the notation:
  NP is the nonsinglet  pseudoscalar and SP is the singlet
  pseudoscalar. This mixing can be estimated from partially quenched
  two flavour QCD~\cite{McNeile:2000hf}.

\begin{figure}
  \begin{center}
  \includegraphics[scale=0.35]{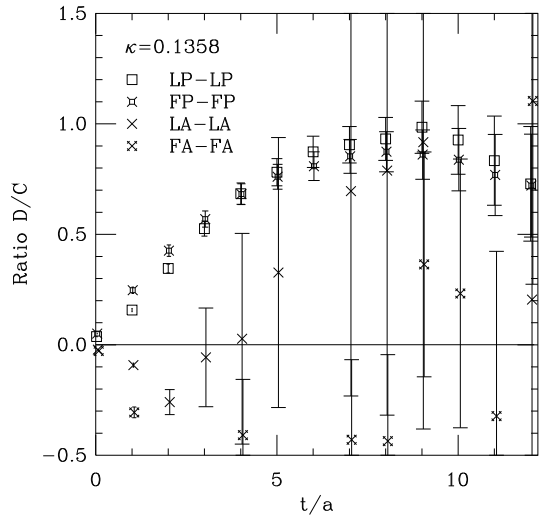}
  \caption[]{The ratio of disconnected $D$ to connected $C$ contributions
  to the SP two-point correlator ($C-D$) at $\kappa = 0.1358$. The operators 
  used for the pion are local (L) of fuzzed (F) and either $\gamma_5$ (P) 
  or $\gamma_5 \gamma_4$ (A). }
  \label{fig_D_by_C}
  \end{center}
  \end{figure}

  We use a similar methodology to that used  in a earlier study on a
  coarser  lattice~\cite{McNeile:2000hf}. The fermion loop was computed
  using complex $Z_2$ noise using the ``two-source'' trick we used in the
  calculation of the  non-singlet scalar~\cite{McNeile:2000xx}. We used 100
  noise samples. Using fuzzed smearing functions as a basis,
  we fitted to a 2 by 2 matrix of correlators using ``factorising fits'' . The
  $A_4$ operator also couples to the pion so we sometimes use a basis of 4
  smearing functions. 
   To show the quality of data we plot the ratio of  disconnected to
  connected correlator in \fig{fig_D_by_C}.  Note that unitarity 
  requires that the SP correlator ($C-D$) is positive so that $D/C < 1$,
  as  we do indeed find. 

  The results from the fits are in \tab{tab:dummy}.
  As expected, the fits with a larger basis of smearing functions
  have smaller
  error bars.

  \begin{table}
  \begin{center}
  \begin{tabular}{|lllllll|} \hline
  $\kappa$ & Corr & region &  $a m_0$  & $a m_1$ & $a m_2$  
  &  $\chi^2/dof$ \\  \hline
  $0.1358$ & a & 2 - 9 &
  $0.497^{+238}_{-137}$ & $1.21^{+35}_{-12}$ & - & 1.5/18 \\
  $0.1358$ & b & 2 - 9 &
  $0.623^{+72}_{-80}$ & $1.965^{+204}_{-225}$ & - & 34/46 \\ \hline
  $0.1355$ & a & 2 - 9 &
  $0.489^{+76}_{-83}$ & $1.45^{+16}_{-16}$ & - & 5.3/18 \\ 
  $0.1355$ & c & 2 - 9 &
  $0.432^{+37}_{-40} $ & $0.75^{+12}_{-12}$ & $2.10^{+25}_{-25}$ &
  48/65 \\ 
  $0.1355$ & b1 & 2 - 9 &
  $0.554^{+40}_{-42} $ & $1.09^{+40}_{-42}$ & $1.98^{+9}_{-9}$ &
  28/41 \\ \hline
  \end{tabular}
  \end{center}
   \caption{Fits to the SP particle. Correlators LL, FL and FF  are used in
  each case with a) having PP only; b) having PP, AP and AA, while  c) has
  all of PP, AP, PA and AA, where $P$ is the pseudoscalar coupling
  ($\gamma_5$)  and A is the time component of the axial ($\gamma_5
  \gamma_4$). The symbol 1 implies momentum=1.}
   \label{tab:dummy}
  \end{table}

   For the fits to the $\kappa=0.1355$ data we could obtain a fit with
  three exponentials. We regard the last  exponential as representing the
  truncation error, hence we have information on one excited  state. The
  mass of the excited state at $\kappa = 0.1355$ is 1.5(2) GeV
  with unknown systematic errors from  the lack of 
  continuum extrapolation and   chiral extrapolation. This is
  encouragingly close to  the the mass of the $\eta(1295)$  and
  $\eta(1440)$ mesons. When the systematics in the lattice calculation 
  are under control, the comparison with experiment will also require an
  understanding of the mixing.

   \begin{figure}
  \begin{center}
  \includegraphics[scale=0.4]{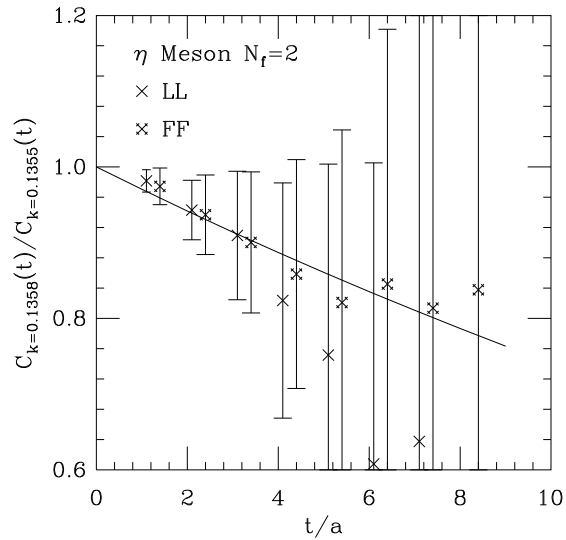}
  \caption[]{The ratio of SP ($N_f=2$ $\eta$ meson) two point 
  correlations from  $\kappa = 0.1358$ to $\kappa = 0.1355$. 
  The curve is described in the text.}
  \label{fig_kratio}
  \end{center}
  \end{figure}

   For the $\kappa=0.1358$ data, we are unable to obtain convincing 
  and consistent fits. Those shown illustrate the problem. We use a
  different method  to show the impact of the data: namely a direct
  comparison between the  SP correlators at the two kappa values, see
  \fig{fig_kratio}. For the LL and FF correlators (with $\gamma_5$ at
   source and sink) the ratios are consistent with  $am_{SP}(0.1358) -
  am_{SP}(0.1355)=-0.03(3)$ as shown, where we have relied more on the  FF
  data since it has a larger contribution from the ground state. This 
  mass difference can then be used to extract an estimate of the SP  mass
  at $\kappa=0.1358$ of $am_{SP}=0.40(5)$.  Note that although the  SP
  mass is approximately constant as the quark mass is reduced, the  pion
  mass (in lattice units) decreases by about 0.09 and thus the difference
  between the  SP mass and the pion mass increases. This large mass
  splitting  is consistent with the steep rise  shown in
  \fig{fig_D_by_C}.

   Because the signal to noise is so poor, we also explore singlet
  correlators with non-zero momentum.   One has to be careful since, at
  non-zero momentum, the axial (A$_4$) pion operator ($g_5g_4$) has
  contamination from $a_1$ --- so factorising fits need additional care  for
  the A$_4$A$_4$ term. For $\kappa=0.1355$,  we can fit for momentum
  $(1,0,0)$ (in units of $2 \pi/L$). For $\kappa=0.1358$, we find  no
  useful additional constraint from the momentum non-zero correlators.

  The mass of the SP meson in two flavour QCD is not immediately available
  from experiment since  the well known mixing between the  $\eta$ and $\eta'$ 
  obscures this issue. By assuming some mixing scheme, we can obtain an 
  estimate of the mass of the SP particle. In our previous
  analysis~\cite{McNeile:2000hf}, the mass of the SP meson was $m_0 =
  \sqrt{ m_{ss}^2 + 2 x_{ss}}$. Using values consistent with phenomenology
   and our previous lattice data ($m_{ss}$ = 0.695 GeV, $x_{ss} = 0.13$
  $GeV^2$), we obtained  0.861 GeV as the mass of the SP meson in  $N_f = 2
  $ QCD.

  Another approach is to use the  Witten-Veneziano expression. The
  SESAM/$T\chi L$  collaboration~\cite{Struckmann:2000bt} obtain 715 MeV
  for mass of the SP in $N_f$ = 2 QCD.

   The chiral extrapolation formulae used  in the lattice QCD
  literature~\cite{Struckmann:2000bt,Lesk:2002gd} have  either used the
  mass or square of the mass  of the SP meson linear in the quark mass. It
  would be clearly better to have a  more theoretical justification of the
   light quark mass dependence, although ``traditional'' 
  chiral perturbation theory 
  is not reliable at mass scales 
  appropriate to the SP state~\cite{Kaiser:2000gs,Borasoy:2001ik}.

\begin{figure}
  \begin{center}
  \includegraphics[scale=0.35]{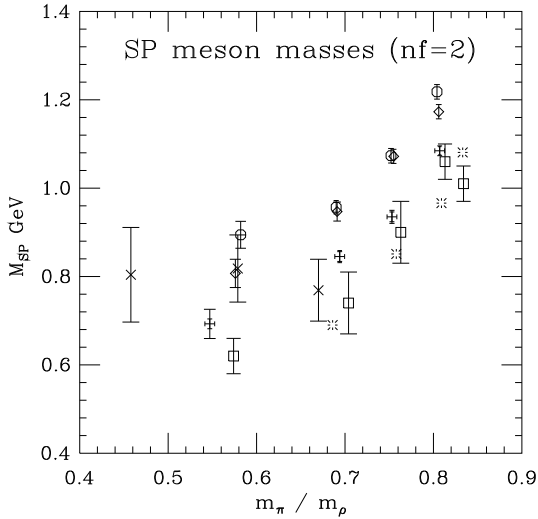}
   \caption[]{World eta data from lattice studies with two degenerate sea
  quarks. The bursts and squares are the values from
  SESAM from the truncated eigenvalue analysis~\cite{Neff:2002mq}
  and Z2 noise measurements~\cite{Struckmann:2000bt} respectively.
   The crosses  are from UKQCD  ($\kappa = 0.1355$ and $\kappa = 0.1358$
  from this work  and $\kappa=0.1398$ from ref.~\cite{McNeile:2000hf}). 
  Results from CP-PACS~\cite{Lesk:2002gd}  are shown by a diamond at  
  $\beta=2.1$, an  octagon at $\beta=1.95$ and  a fancy plus at
  $\beta=1.8$
   }
  \label{WorldEta}
  \end{center}
  \end{figure}

  In \fig{WorldEta}, the world data for  the SP mass is plotted as a
  function of the   pseudoscalar to vector  mass ratio. Our point at 
  $\kappa = 0.1355$ is consistent with the data from other groups. The
  mass at $\kappa = 0.1358$ lies above the trend from larger quark mass,
  but  we do expect the SP mass to go to a non-zero constant as the quark
  mass vanishes. Indeed our semi-phenomenological estimate given  above
  was that the SP mass in the chiral limit is  $861$ MeV. This value is very
  consistent with the flattening  behaviour indicated by our data point.

  For their final result, CP-PACS~\cite{Lesk:2002gd} quoted  the mass of
  the SP particle as $0.960(87)_{-0.248}^{+0.036}$ GeV. This high value
  arises  from  the continuum extrapolation. The central value
  for the mass was obtained by linearly extrapolated in lattice spacing
  with a  $\chi^2/{\mathrm{dof}}$ of $4.2$. 
  A quadratic extrapolation  in the lattice
  spacing had a $\chi^2/{\mathrm{dof}}$  of  $2.8$, with the resulting mass of
  $0.819(50)$ GeV. The use of a linear extrapolation with lattice spacing is
  consistent with the rest of the spectroscopy program of
  CP-PACS~\cite{AliKhan:2001tx}. The large errors in the mass of the SP
  meson from CP-PACS represent the variation from the different continuum
  extrapolations.  

  \section{Topology}
  \label{sec:top}

  \begin{figure}

  \begin{center}
  \includegraphics[width=\figwidth,clip]{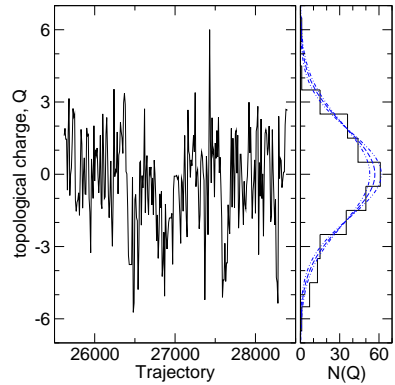}

  \caption[]{A time history of the topological charge, $Q$, with a
  histogram using unit--sized bins, and the Gaussian curve derived
  from the central value of topological susceptibility (and one standard
  deviation either side as outlying curves).
  \label{fig_q}
  }
  \end{center}
  \end{figure}

  The topological charge, $Q$, and its associated susceptibility,
  \begin{equation}
  \chi = \frac{\langle Q^2 \rangle}{L^3 T} \; ,
  \end{equation}
  are expected to be especially sensitive to the presence and 
  properties of the sea quarks in QCD. 
  As the mass of the sea quarks is reduced (towards the chiral
  limit), the topological susceptibility is suppressed below its
  quenched value. For sufficiently large volumes, the leading order
  chiral behaviour is
  \cite{Crewther:1977ce,DiVecchia:1980ve,Leutwyler:1992yt}
  \begin{equation}
  \chi(m_\pi^2) = \frac{(f_\pi m_\pi)^2}{4} + {\cal O}(m_\pi)^4
  \label{eqn_chi_chi}
  \end{equation}
  for two degenerate flavours (and using a normalisation where $f_\pi
  \simeq 93 ~ \mev$). As $m_q \to \infty$, $\chi \to \chi^{\text{qu}}$ the
  quenched value which is around $(180 \MeV)^4$.  The higher order
  corrections in \eq{eqn_chi_chi} must therefore introduce a negative
  curvature at some intermediate quark mass
  \cite{Leutwyler:1992yt,Hart:2000wr,Durr:2001ty}.

  On the lattice, the topological susceptibility becomes renormalised
  relative to the continuum value, $\chi_{\text{cont}}$, both
  multiplicatively and additively
  \cite{Bernard:2003gq}:
  \begin{equation}
  \chi = Z^2 \chi_{\text{cont}} + M \; .
  \label{eqn_chi_renorm}
  \end{equation}
  Broadly speaking, $M \ge 0$ arises from the presence of
  ``dislocations'': short range fluctuations in the gauge field that
  masquerade as small instantons. $Z \le 1$ reflects the breaking of
  scale invariance on the lattice, whereby small instantons have a
  topological charge less than unity.  At large quark masses (or in the
  quenched theory) the first term in \eq{eqn_chi_renorm} dominates,
  suppressing $\chi$ at non--zero $a$. In the chiral limit, however, $M$
  dominates and is non--zero even after smoothing. The topological
  susceptibility then shows strong discretisation effects that act to
  \textit{increase} $\chi$ at finite lattice spacing.

  Whilst comparing ``matched'' ensembles at fixed lattice spacing will
  control discretisation effects, these opposite trends imply that they
  will not cancel away entirely. The net effect is that is that any
  chiral suppression of the topological susceptibility at given $a$
  relative to the quenched value at an equivalent lattice spacing will
  be less than in the continuum limit.

  We measure $Q$ using the method of
  \cite{Hart:2000wr}:
  $n_c=10$ cooling sweeps are applied using the
  Wilson gauge action. Ten cools strikes a good balance between
  adequate suppression of these ultraviolet dislocations and excessive
  destruction of the long range topological structure
  \cite{Hart:2000wr}.
  A reflection--symmetrised ``twisted plaquette'' lattice topological
  charge operator is used
  \cite{Hart:1999hy}.

\begin{figure}

  \begin{center}
  \includegraphics[width=\figwidth,clip]{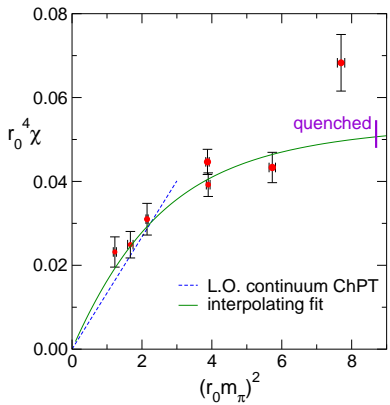}

  \caption[]{The topological susceptibility as a function of the
  lightest pseudoscalar (``pion'', $\kappa_{\mathrm{valence}} =
  \kappa_{\mathrm{sea}}$) mass for two flavours of ${\cal
  O}(a)$--improved fermions. Quenched values for this range of lattice
  spacing are shown as a bar on the RHS of the plot.
  \label{fig_khi}
  }
  \end{center}
  \end{figure}

  As discussed previously, good decorrelation of $Q$ is seen in the
  simulation, and the histogram of the populations of the different
  topological sectors in \fig{fig_q} has the expected Gaussian
  form. We find $\langle Q \rangle = -0.33 (29)$, consistent with
  zero. The susceptibility is $\chi = 0.292 (45) \times
  10^{-4}$, or $0.284 (34) \times 10^{-4}$ if we subtract terms in
  $\langle Q \rangle$. We plot the latter result as the leftmost point
  in \fig{fig_khi}, alongside previously published UKQCD results 
  \cite{Hart:2001tv,Hart:2001pj,Allton:2001sk,Hart:2000gh,
  Hart:2000wr,Hart:1999hy}.
  We show also the equivalent quenched susceptibility as a bar whose
  length reflects both the statistical uncertainty in the quenched
  measurements and the small variation in lattice spacing across the
  ensembles depicted, despite the presence of discretisation effects. We
  see very clear evidence for a strong chiral suppression of $\chi$
  relative to the quenched theory, driven by the sea quarks.

  For \eq{eqn_chi_chi} to hold, we require 
  $x_{\text{LS}} \equiv (f_\pi m_\pi)^2 V \gg 1$
  \cite{Leutwyler:1992yt}.
  Using the continuum value of $f_\pi$ (which is lower than the value at
  finite lattice spacing) we find $x_{\mathrm{LS}} = 11$. Finite volume
  effects appear to be less significant for $\chi$ than, for instance,
  the light hadron spectrum. Studies for this action indicate that
  $x_{\mathrm{LS}} \gtrsim 10$ is sufficient for such finite size
  effects to be within the statistical uncertainty in $\chi$
  \cite{Hart:2004ij}.

  As discussed above, comparison of data at finite lattice spacings with
  continuum predictions must be made cautiously. That being said,
  Fig.~\ref{fig_khi} is very encouraging and is evidence for the
  improved chiral properties of the ${\cal O}(a)$--improved
  action. Performing a leading order [in $(r_0 m_\pi)^2$] fit, or an
  interpolating fit across the chiral range
  \cite{Hart:2001pj},
  we see that the slope near the origin of Fig.~\ref{fig_khi} is
  slightly greater than that expected from the continuum value of
  $f_\pi$. As discretisation effects are expected to increase the
  pseudoscalar decay constant, this is in agreement with theoretical
  expectations. In a forthcoming paper discretisation effects will be
  examined in more detail
  \cite{Hart:2004ij}.

  Finally, it is interesting to compare these results with those
  obtained for three flavours of improved staggered sea quarks
  \cite{Bernard:2003gq}.
  For similar lattice spacings, $a \simeq 0.09 \fm$, the topological
  susceptibility was non--zero and roughly constant below $(r_0 m_\pi)^2
  \simeq 2$, when $\chi$ presumably became dominated by $M$. No
  statistically significant evidence for such a trend can yet be seen
  for the ${\cal O}(a)$--improved action, however.

  \section{Conclusions}
  \label{sec:conc}

  Although we are excited to have finally reached
  a region of parameter space where unquenched clover 
  calculations are starting to see chiral logs and 
  uncover interesting behaviour in the mass of the singlet 
  pseudoscalar meson, it is not yet clear how to
  improve on these results. 

  If we implemented some of the new updating
  algorithms~\cite{Hasenbusch:2002ai,Hasenbusch:2003vg}
  for clover fermions, in principle we could work
  at a larger volume with fewer of the problems
  reported in section~\ref{sec:simul}. However, concerns 
  have been raised about
  the interaction~\cite{Jansen:2003nt,Sommer:2003ne} 
  of the Wilson gauge action with 
  the clover fermion action in unquenched calculations.
  Once improved gauge actions have been incorporated into
  the non-perturbative clover improvement formalism
  for fermions, we could study the chiral log structure,
  at fixed lattice spacing~\cite{Aoki:2003yv,Bar:2003mh},
  This would allow important comparisons with the results
  from other unquenched lattice QCD calculations that use
  different fermion formalisms~\cite{Davies:2003ik}.

  \begin{acknowledgments}

  We acknowledge the support of the U.K. Particle Physics and Astronomy
  Research Council under grants GR/L22744, GR/L29927, GR/L56374,
  PPA/G/O/1998/00621, PPA/G/O/1998/00518, PPA/G/S/1999/00532,
  PPA/G/O/2000/00465 and PPA/G/O/2001/00019. 
  A.H. is supported by the
  U.K. Royal Society.
  We thank EPCC for time on Lomond. 
  We are grateful to the ULgrid project of the
  University of Liverpool for computer time.
  The authors acknowledge support from EU
  grant HPRN-CT-2000-00145 Hadrons/LatticeQCD.

  \end{acknowledgments}

  %
  %
  %
  %
  %

\end{document}